\documentclass[mathleft]{an}
\usepackage{graphicx}
\usepackage{times}
\begin{document}
\sloppy
 \Pagespan{1150}{}
 \Yearpublication{2007}
 \Yearsubmission{2007}
 \Month{9}%
 \Volume{328}%
 \Issue{10}%
 \DOI{10.1002/asna.200710859}%

\title{Stability of toroidal magnetic fields in the solar tachocline and beneath}

\author{L.L.~Kitchatinov\inst{1,2}\fnmsep\thanks{Correspondence to:
  {kit@iszf.irk.ru}}
  \and G.~R\"udiger\inst{1} 
}
 \titlerunning{Stability of toroidal magnetic fields}
 \authorrunning{  L.L.~Kitchatinov \& G.~R\"udiger}
 \institute{ Astrophysikalisches Institut Potsdam, An der Sternwarte
             16, D-14482 Potsdam, Germany
 \and Institute for Solar-Terrestrial Physics, P.O.~Box 291, 
      Irkutsk 664033, Russia}

 \received{2007 Sep 25}
 \accepted{2007 Oct 22}
 \publonline{2007 Dec 15}


 \abstract{%
Stability of toroidal magnetic field in a stellar radiation zone is considered  for the cases of uniform and differential rotation. In the rigidly rotating radiative core shortly below the tachocline, the critical magnetic field for instability is about 600~G. The unstable disturbances for slightly supercritical fields have short radial scales $\sim$1~Mm. Radial mixing produced by the instability is estimated to conclude that the internal field of the sun can exceed the critical value of 600~G only marginally. Otherwise, the mixing is too strong and not compatible with the observed lithium abundance. Analysis of joint instability of differential rotation and toroidal field leads to the conclusion that axisymmetric models of the laminar solar tachocline are stable to nonaxisymmetric disturbances. The question of whether sun-like stars can posses tachoclines is addressed with positive answer for stars with rotation periods shorter than about two months. 
 }

\maketitle

\section{Introduction}\label{introduction}
The problem of stability of toroidal magnetic fields in stellar interiors first
addressed by  Tayler (\cite{T73}) is highly relevant for the dynamics of stellar radiation zones. The stability controls which primordial fields can survive inside the stars on evolutionary
time scales. Magnetic instabilities are important for transport of angular
momentum and chemical species in the radiation zones (Barnes, Charbonneau \&
MacGregor \cite{BCM99}). The stability problem is also relevant to the origin of the solar tachocline (Gilman \cite{G05}).

This paper concerns the stability of toroidal magnetic fields in rotating
radiation zones. The analysis is global in horizontal dimensions but the radial
scales of unstable disturbances are assumed as short. The smallness of the
radial scales is a  consequence of the stable stratification. Our mathematical
formulation is close to the thin layer approximation of Cally (\cite{C03}) and
Miesch \& Gilman (\cite{MG04}). Our aims, however, are quite different. We do
not address the growth rates of highly supercritical disturbances but focus on
the marginal stability. We also include finite diffusion that is necessary when
stability of rigidly rotating region beneath the tachocline is concerned. When
diffusion is neglected, the most rapidly growing  disturbances have indefinitely
short radial scales and the instability can be excited by indefinitely small
magnetic field. The unphysical zero wave lengths and field amplitudes are
prevented by finite diffusion. The marginal field strength for the rigidly
rotating region just beneath the tachocline is about 600~G. Estimation of the
radial mixing produced by the instability suggests that the marginal value is
close to the upper bound for the internal field. The mixing is too strong and
is not compatible with the observed lithium abundance if the marginal value is
considerably exceeded.

Differential rotation can drive its own instability even without magnetic fields (Watson \cite{W81}; Dziembowski \& Ko\-so\-vichev \cite{DK87}). The instability has been extensively studied in 2D approximation of strictly horizontal disturbances. An account for radial displacements in our 3D formulation reduces the magnitude of differential rotation required for the hydrodynamic instability. 

We find  that axisymmetric models of a laminar solar tachocline do not suffer
from nonaxisymmetric magnetic instabilities. The question arises whether  stars other than the
Sun can possess tachoclines. Evidences on stellar differential rotation used in
combination with laminar tachocline theory provide a positive answer for all solar-like stars whose rotation is not too slow.
\section{The model}\label{model}
The background state of our model includes differential rotation and toroidal magnetic field. The angular velocity distribution is parameterized as
\begin{equation}
   \Omega = \Omega_0\left(1 - a\cos^2\theta\right) ,
   \label{1}
\end{equation}
where $\Omega_0$ is the equatorial value of $\Omega$ and $\theta$ is 
colatitude. The toroidal field is expressed in terms of the Alfv\'en
angular frequency $\Omega_\mathrm{A}$, i.e.
\begin{equation}
   {\vec B} =  r\sin\theta\sqrt{\mu_0\rho}\
   \Omega_\mathrm{A}\left( r,\theta\right) {\vec e}_\phi,
   \label{2}
\end{equation}
where $\vec{e}_\phi$ is the azimuthal unit vector. The field is
assumed to consist of two broad belts antisymmetric about the solar
equator:
\begin{equation}
   \Omega_\mathrm{A} = \Omega_0 b \cos\theta .
   \label{3}
\end{equation}
$b$ is the normalized field strength, $b=1$ means approximate
equipartition of rotational and magnetic energies.

\begin{figure}
    \includegraphics[width=8.cm, height=4.7cm]{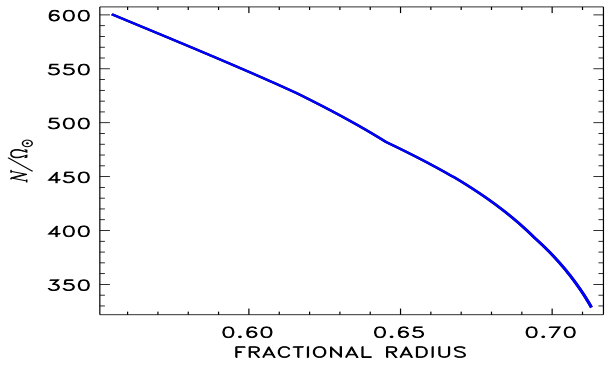}
    \caption{(online colour at: www.an-journal.org) The buoyancy frequency 
            (Eq. \ref{4}) in the upper part
            of the radiative core of the Sun after the 
            model by Stix \& Skaley (\cite{SS90}). The convection zone 
            of the model includes the overshoot layer
            so that $N/\Omega_\odot$ is large immediately beneath
            the convection zone.
              }
    \label{f1}
\end{figure}

The equations for linear perturbations written in the Appendix were derived in
Kitchatinov \& R\"udiger (\cite{KR07}). The equations apply to the disturbances
that are global in horizontal dimensions but short-scaled in radius. The reason
why the radial scales are short is the stable (subadiabatic) stratification. The buoyancy frequency, $N$,
\begin{equation}
   N^2 = \frac{g}{C_\mathrm{p}}\frac{\partial s}{\partial r} ,
   \label{4}
\end{equation}
enters the normalized equations  via the parameter
\begin{equation}
   \hat\lambda = \frac{N}{\Omega_0 k r} ,
   \label{5}
\end{equation}
where $g$ is the gravity, $C_\mathrm{p}$ is the heat capacity at constant
pressure, $s = C_\mathrm{v}\ln\left(P/\rho^\gamma\right)$ is the specific entropy, $r$ is
the (heliocentric) radius and $k$ is the radial wave number of the disturbances.
The ratio $N/\Omega_0$ in stellar radiation zones is a large quantity
(Fig.~\ref{f1}). After Eq.~(\ref{5}), disturbances can avoid the stabilizing
effect of stratification by decreasing their radial scales. Our results shall
confirm   this short-wave approximation with $kr \ll 1$.

The equations for small disturbances listed in the Appendix allow two types of
equatorial symmetry. We use the standard notations Sm and Am for symmetric and
antisymmetric modes, $m$ is the azimuthal wave number.  Sm modes have
mirror-sym\-met\-ry (about the equatorial plane) disturbances of the magnetic
field ${\vec B}'$ (symmetric $B'_\phi$, $B'_r$ and antisymmetric $B'_\theta$),
mirror-an\-ti\-sym\-met\-ric velocity ${\vec u}'$ (symmetric $u'_\theta$ and
anti\-sym\-met\-ric $u'_\phi$ and $u'_r$) and antisymmetric perturbations of
entropy $s'$. The symmetries reverse for Am modes. These symmetry conventions
are the same as in Gilman, Dikpati \& Miesch (\cite{GDM07}).

\section{Results and discussion}
\subsection{Beneath the tachocline ($\vec{a}$\ =\ 0)}
Helioseismology shows that the rotation below the tachocline is almost rigid
(Schou et al. \cite{Sea98}). The uniform rotation has minimum kinetic energy for
given angular momentum. Rotational energy cannot, therefore, feed any
instability so that  magnetic instabilities can only be considered. 

\subsubsection{Ideal fluids}
\begin{figure}[!tb]
   \includegraphics[width=7.0cm,height=4.7cm]{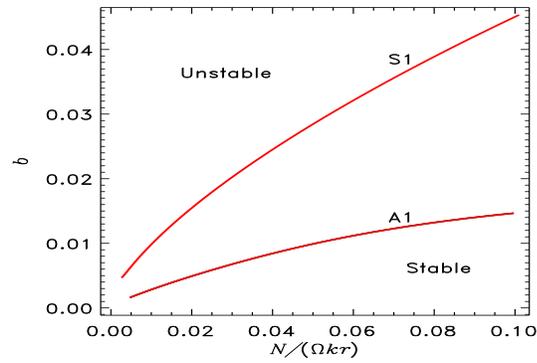}
   \caption{(online colour at: www.an-journal.org) Stability map for ideal fluid  of zero diffusion and rigid
            rotation. Indefinitely small radial scales are
            preferred.
              }
   \label{f2}
\end{figure}

Microscopic diffusion in radiative cores is small and often neglected. We start with the case of ideal fluids with diffusion parameters  equal zero. The stability map for this case is shown in Fig.~\ref{f2}.

Only the kink-type ($m=1$) disturbances are found unstable. The instability remains
active for indefinitely small field amplitude $b$  though the range of radial
scales of unstable excitations reduces with decreasing $b$. Therefore, rotation
does here not suppress the Tayler instability. The character of rotational influence
crucially depends on the toroidal field profile. Pitts \& Tayler (\cite{PT85}) found strong rotational suppression but suggested that it might be restricted to the case in which the ratio of rotational to Alfv\'en velocities is uniform. The results of Fig.~\ref{f2} confirm the expectation. The results were found with same numerical code as formerly reported strong rotational quenching for uniform $\Omega_\mathrm{A}$ (Kitchatinov \& R\"udiger \cite{KR07}).

The ideal instability for very weak fields is, however, of mainly academic interest because the unstable disturbances have indefinitely small radial scales. Finite diffusion must, therefore, be included.

\subsubsection{Finite diffusion}
From now on  the finite values
\begin{equation}
 \epsilon_\nu = 2\cdot 10^{-10}, \ \ \ \ \ \
 \epsilon_\eta = 4\cdot 10^{-8}, \ \ \ \ \ \
 \epsilon_\chi = 10^{-4},
   \label{6}
\end{equation}
that are characteristic of the upper part of the solar radiative
core are used for the diffusion parameters (Eq. \ref{A6}). The relation $\chi\gg\eta\gg\nu$ of Eq.~(\ref{6}) is typical of stellar radiation zones.

\begin{figure}[tb]
   \includegraphics[width=7.0cm,height=5.0cm]{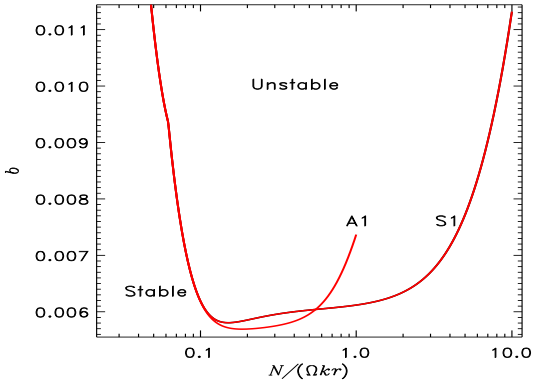}
   \caption{(online colour at: www.an-journal.org) Stability map for finite diffusion and rigid rotation.
              }
   \label{f3}
\end{figure}

The stability map computed with finite diffusion is shown in Fig.~\ref{f3}. Small radial scales are now stable due to finite magnetic diffusivity.

\begin{figure}[!tb]
   \centering{
   \includegraphics[width=4.cm,height=4.5cm]{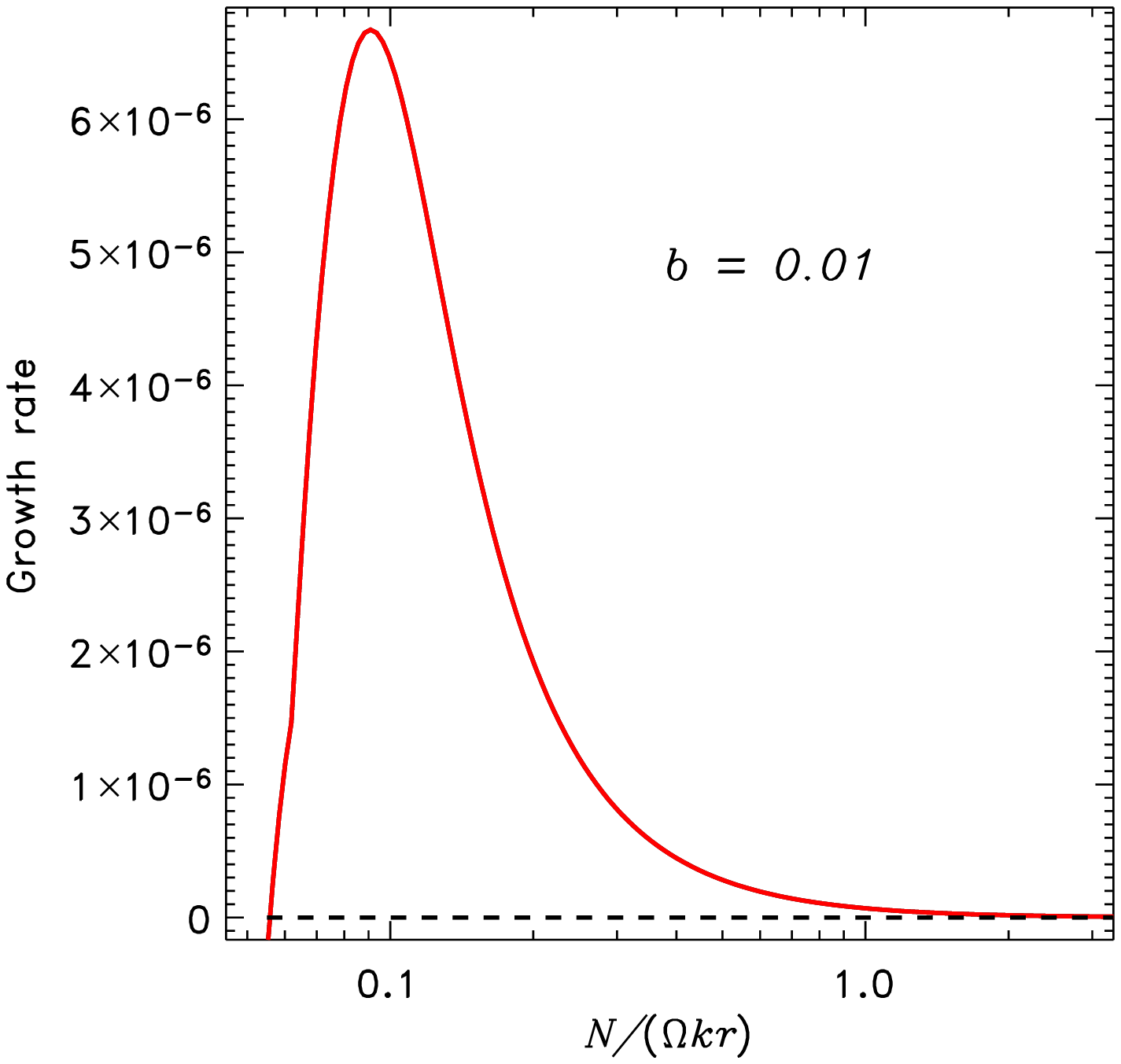}
   \hspace{0.2cm}
   \includegraphics[width=4.cm,height=4.5cm]{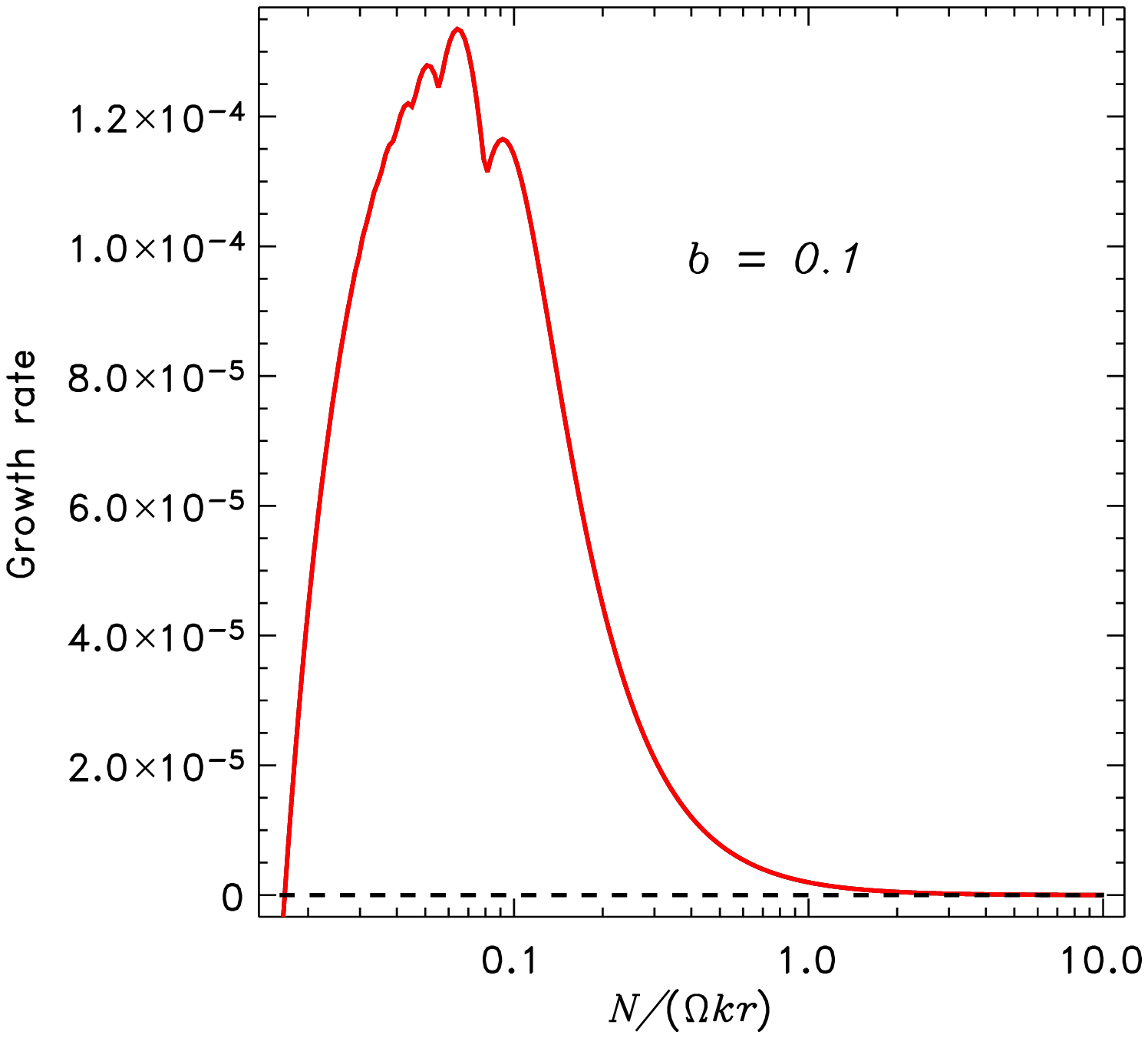}}
   \caption{(online colour at: www.an-journal.org) Growth rates in units of $\Omega_0$  
        for $b=0.01$ (\emph{left}) and 
        $b=0.1$ (\emph{right}).
        The largest rates belong to ${\hat\lambda \;\buildrel
        <\over{\scriptstyle\sim}\; 0.1}$. All curves for most rapidly growing modes.
              }
   \label{f4}
\end{figure}

For the upper radiation zone of the Sun with $\rho\simeq 0.2$~g/cm$^3$ it follows 
\begin{equation}
    B_\phi \simeq 10^5 b \ \ \ \mathrm{G}, 
    \ \ \ \ \ \ \ \
     \lambda \simeq 10\hat\lambda\ \ \ \mathrm{Mm} .
     \label{8}
\end{equation}
After Fig.~\ref{f3}, the critical magnetic field for the onset of the
instability is thus about 600~G. The modes that first become unstable when this
field strength is exceeded have vertical wavelengths between 1 and 2 Mm. For
larger field strengths, there is a range of unstable wavelengths. The maximum
growth rates remain, however, at the wavelengths $\sim 1$~Mm (Fig.~\ref{f4}).
The growth rates are rather small.

\begin{figure}[tb]
   \includegraphics[width=7.0cm, height=5.0cm]{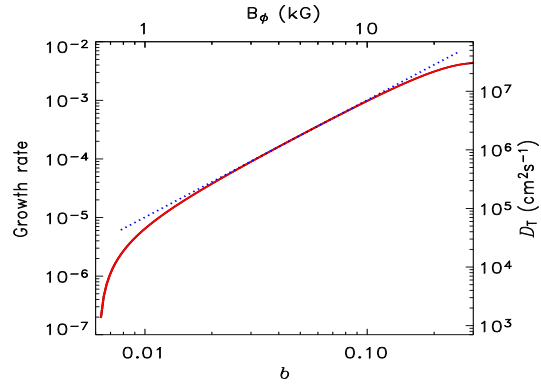}
   \caption{(online colour at: www.an-journal.org) The growth rate $\hat\sigma$
            in units of $\Omega_0$ as function of the toroidal field
            amplitude.
            The dotted line shows the parabolic approximation
            $\hat\sigma = 0.1 b^2$. The scale
            on the right gives the radial diffusivity
            of chemical species estimated after Eq. (\ref{9}).
              }
   \label{f5}
\end{figure}

The instability radially mixes chemicals. It can thus be relevant for the radial
transport of light elements (Barnes et al. \cite{BCM99}). The effective
diffusivity, $D_\mathrm{T} \simeq \lambda^2\sigma$  can be estimated from our
linear computations.
With Eq.~(\ref{8}) this yields
\begin{equation}
    D_\mathrm{T} \simeq  10^{12} \hat\lambda^2\ \hat\sigma\ \ \
    \mathrm{cm}^2/\mathrm{s},
    \label{9}
\end{equation}
where $\hat\sigma$ is the normalized growth rate given in
Fig.~\ref{f5}. Diffusivities in excess of $10^4$\,cm$^2$s$^{-1}$ in
the upper radiative core are not compatible with the observed solar
lithium abundance. Hence, the toroidal field amplitude can only
slightly exceed the marginal value of about 600~G. 

The estimation of light elements mixing seem to exclude the possibility 
of hydromagnetic dynamo driven by magnetic instability in the solar radiation
zone (Spruit \cite{S02}).
This possibility is also not supported by direct numerical simulations of 
Zahn, Brun \& Mathis (\cite{ZBM07}). 

\subsection{Inside the tachocline ($\vec{a>}$  0)}
The helioseismology detections of the tachocline central radius
$r_\mathrm{c} = (0.693\pm 0.002)\,{\rm R}_\odot$ and (equatorial) thickness
$\Delta r = (0.039 \pm 0.013)\,{\rm R}_\odot$ (Charbonneau et al.
\cite{Cea99}) place the tachocline mainly if not entirely  
beneath the base of convection zone at $r = 0.713\,{\rm R}_\odot$
(Basu \& Antia \cite{BA97}; Christensen-Dalsgaard, Gough \& Thompson
\cite{C-Dea91}). The stability analysis of toroidal fields of
radiative core can thus be applied to the tachocline. The
differential rotation with $a>0$ in Eq.~(\ref{1}) must also be included.

\subsubsection{Hydrodynamic stability}\label{Watson}
Differential rotation can be unstable even without magnetic fields. The hydrodynamic stability of the tachocline 
has been studied in 2D approximation of strictly horizontal displacements 
(Charbonneau, Dikpati \& Gilman \cite {CDG99}; Garaud \cite{G01}). Here we extend the treatment to 3D.

\begin{figure}[htb]
   \includegraphics[width=7.0cm, height=5.0cm]{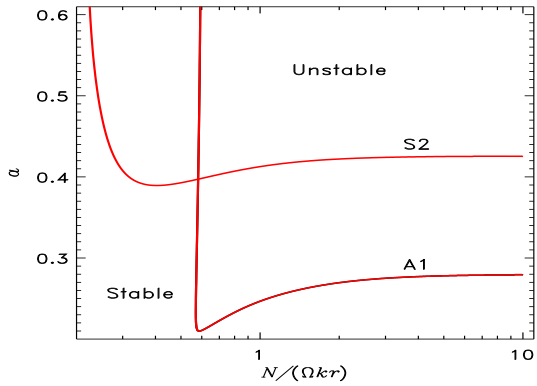}
   \caption{(online colour at: www.an-journal.org) The map of hydrodynamic instability of
            differential rotation (\ref{1}). Most unstable are the
            perturbations of A1 symmetry type with the vertical scale
            $\hat\lambda \simeq 0.6$. The critical magnitude of latitudinal
            shear is reduced to 0.21 compared to the 0.28 of 2D
            theory.
              }
   \label{f6}
\end{figure}

The resulting stability map is shown in Fig.~\ref{f6}. A1 and S2 modes can be
unstable for positive $a$. The marginal values of $a$ approach constants for
large $\hat\lambda$. The large-$\hat\lambda$ limit reproduces the results of the
2D theory as it should  (Kitchatinov \& R\"udiger \cite{KR07}). The minimum
$a$-value for the instability corresponds, however, to finite vertical scales of about 6 Mm. 

\subsubsection{Joint instabilities}
Growth rates of joint instabilities for $b\sim 1$ have been discussed by Cally
(\cite{C03}) and Gilman et al. (\cite{GDM07}). Now we concern the marginal stability that is important in relation to the models of laminar tachocline formed by a weak internal magnetic field (R\"udiger \& Kitchatinov \cite{RK07}).

\begin{figure}[tb]
   \includegraphics[width=7.0cm,height=5.0cm]{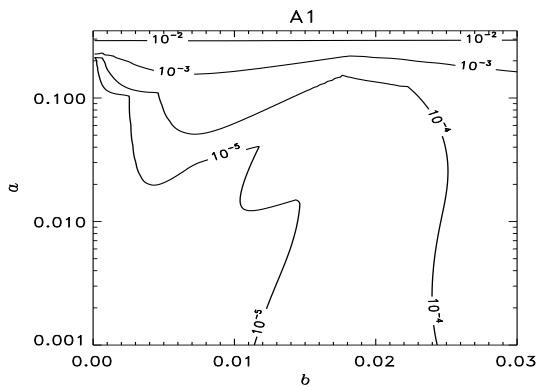}
   \caption{Isolines of small growth rates on the plane of differential rotation
            and toroidal field parameters $a$ and $b$ of
            Eqs.~(\ref{1}) and (\ref{3}). Laminar tachocline models are 
            stable.
              }
   \label{f7}
\end{figure}
\begin{figure}[!b]
   \includegraphics[width=7.0cm,height=5.0cm]{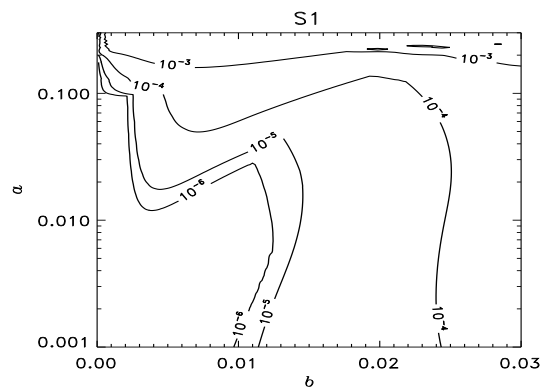}
   \caption{Same as in Fig.~\ref{f7} but for S1 modes. 
              }
   \label{f8}
\end{figure}

Isolines of the very small growth rates for A1 and S1 modes for the $a$ and $b$
parameters of  differential rotation and magnetic field are shown in 
Figs.~\ref{f7} and \ref{f8}. The growth rates are optimized in $\hat\lambda$, i.e. 
$\hat\lambda$  has been varied to find the maximum growth rates for given $a$
and $b$. From Sect.~\ref{Watson} we know that differential rotation is stable to
S1 disturbances. For sufficiently large $a$, the S1 modes become unstable for
very small magnetic field (Fig.~\ref{f8}). The characteristic growth rates of
this mode are nevertheless large compared to S2 disturbances.

The models for the laminar solar tachocline are valid for microscopic diffusivities 
or sufficiently low turbulent diffusivities up to about $10^7$\,cm$^2$/s (Kitchatinov \& R\"udiger
\cite{KR06}). According to Eq. (\ref{9}), this means growth rates up to ${\hat\sigma 
\sim 10^{-3}}$. The laminar tachocline models have toroidal field amplitudes of about 
200~G or $b \simeq 0.002$. The maximum field in the models is attained at 
those depths where latitudinal differential rotation is reduced to about 10\%.
Figures \ref{f7} and \ref{f8} show that the axisymmetric models for the laminar
solar tachocline are  {\em stable to nonaxisymmetric disturbances}.

\subsubsection{Do stars other than the Sun possess tachoclines?}
\begin{figure}[b]
   \includegraphics[width=7.0cm, height=5.0cm]{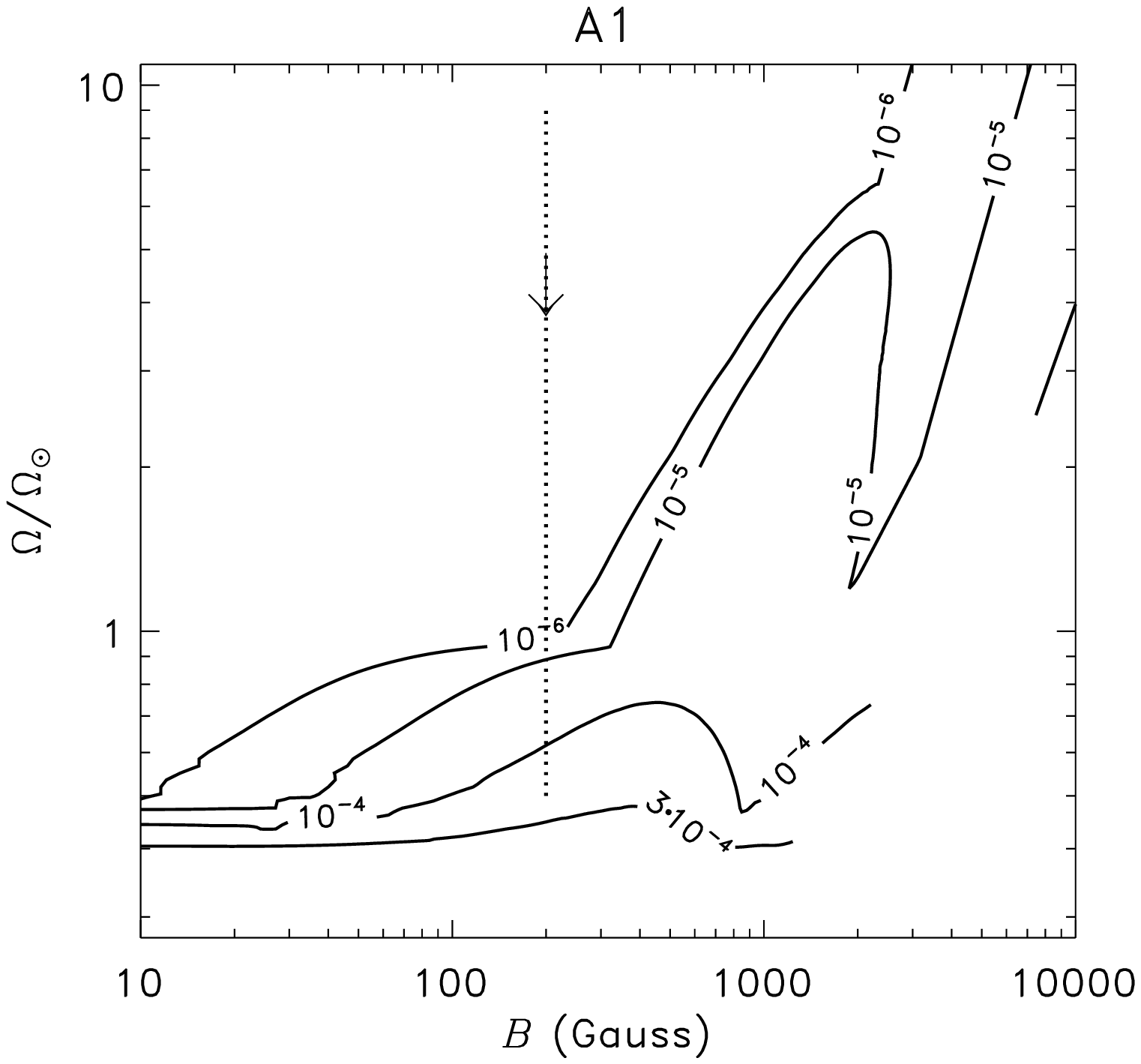}
   \caption{Isolines of small growth rates of A1 modes for various rotation 
            rates and  toroidal field amplitudes. The vertical dashed line 
            shows the evolutionary track of a star according to 
            the laminar tachocline models. Solar mass stars rotating not much 
            slower than the Sun are predicted to possess tachoclines.
              }
   \label{f9}
\end{figure}

Observations of Barnes et al. (\cite{Bea05}) and theoretical models of Kitchatinov \& 
R\"udiger (\cite{KR99}) and K\"uker \& Stix (2001) suggest that the
absolute value $\Delta\Omega$ of differential rotation is almost independent of
the rotation rate for stars of given mass (but increases with mass).
Figure~\ref{f9} reformulates the stability map for the (most easily excited) A1
modes in terms of the magnetic field strength and angular velocity of a star of solar mass assuming $\Delta\Omega = a\Omega_0 = \mathrm{const}$ and using Eqs.~(\ref{2}) and (\ref{3}) to recover $B_\phi$.

The toroidal field amplitude in the laminar tachocline models is controlled by 
$\Delta\Omega$ and fluid density only. Therefore, the limit $B \simeq 200$~G of
the solar tachocline model does also apply to the stars of same mass but
different rotation rates. The star moves down on Fig.~\ref{f9} along a vertical
line of $B \simeq 200$~G as the star spins down with age. The toroidal field can
be only mildly unstable ($\hat\sigma < 10^{-3}$) for the laminar tachocline
model to apply. All solar mass stars with rotation periods shorter than about
two months should thus possess tachoclines after Fig.~\ref{f9}.

\acknowledgements This work was supported by the Deutsche
Fo\-r\-schungs\-ge\-mein\-schaft and by the Russian Foundation for
Basic Research (project 05-02-16326).

\appendix
\section{Linear stability equations}
The equations used in our stability analysis represent the eigenvalue problem
resulting from the exponential time dependence $\exp(-\mathrm{i}\omega t)$ of
the disturbances.  The dependencies on azimuth and radius are taken as Fourier
modes $\exp (\mathrm{i}m\phi + \mathrm{i}kr)$. The equations are formulated in
terms of scalar potentials for the vector magnetic and velocity fields (Kitchatinov \& R\"udiger \cite{KR07}).

The equation for the potential $V$ of the poloidal flow reads
\begin{eqnarray}
   \lefteqn{\hat\omega\left(\hat{\cal L}V\right) =
   -\hat{\lambda}^2\left(\hat{\cal L}S\right)
   - \mathrm{i}\ \frac{\epsilon_\nu}{\hat{\lambda}^2}\left(\hat{\cal L}V\right)}
   \nonumber \\
   &&-\ 2\mu\hat\Omega\left(\hat{\cal L}W\right)
   - 2\left(1-\mu^2\right)\frac{\partial\left(\mu\hat\Omega\right)}
   {\partial\mu}\frac{\partial W}{\partial\mu}
   -2 m^2\frac{\partial\hat\Omega}{\partial\mu} W
   \nonumber \\
   &&+\ 2\mu\hat\Omega_\mathrm{A}\left(\hat{\cal L}B\right)
   + 2\left(1-\mu^2\right)\frac{\partial
   \left(\mu\hat\Omega_\mathrm{A}\right)}
   {\partial\mu}\frac{\partial B}{\partial\mu}
   \nonumber \\
   &&+\ 2 m^2\frac{\partial\hat\Omega_\mathrm{A}}{\partial\mu} B
   - m\hat\Omega_\mathrm{A}\left(\hat{\cal L}A\right)
   - 2m\frac{\partial\left(\mu\hat\Omega_\mathrm{A}\right)}
   {\partial\mu} A
   \nonumber \\
   &&-\ 2m\left(1-\mu^2\right)\frac{\partial\hat\Omega_\mathrm{A}}
   {\partial\mu} \frac{\partial A}{\partial\mu}
   + m\hat\Omega\left(\hat{\cal L}V\right)
   \nonumber \\
   &&+\ 2m\frac{\partial\left(\mu\hat\Omega\right)}{\partial\mu} V
   + 2m\left(1-\mu^2\right)\frac{\partial\hat\Omega}{\partial\mu}
   \frac{\partial V}{\partial\mu},
   \label{A1}
\end{eqnarray}
where $\hat\lambda$ is the normalized radial wave length (\ref{5}).
The dimensionless eigenvalue $\hat\omega$, the angular velocity
$\hat\Omega$ and the Alfv\'en angular frequency
$\hat\Omega_\mathrm{A}$ are all normalized to $\Omega_0$, $\mu =
\cos\theta$, and
\begin{eqnarray}
   \hat{\cal L} = \frac{\partial}{\partial\mu}\left( 1-\mu^2\right)
   \frac{\partial}{\partial\mu} - \frac{m^2}{1 - \mu^2} .
   \nonumber
\end{eqnarray}
The complete system of five equations also includes the equation for the
potential $W$ of toroidal flow
\begin{eqnarray}
  \lefteqn{\hat\omega\left(\hat{\cal L}W\right) =
  - \mathrm{i}\, \frac{\epsilon_\nu}{\hat{\lambda}^2}\left(\hat{\cal L}W\right)
  + m\hat\Omega\left(\hat{\cal L}W\right)}
  \nonumber \\
  &&-\ m\hat\Omega_\mathrm{A}\left(\hat{\cal L}B\right)
  -  mW\frac{\partial^2}{\partial\mu^2}
  \left(\left(1-\mu^2\right)\hat\Omega\right)
  \nonumber \\
  &&+\ mB\frac{\partial^2}{\partial\mu^2}
  \left(\left(1-\mu^2\right)\hat\Omega_\mathrm{A}\right)
  + \left(\hat{\cal L}V\right)\frac{\partial}{\partial\mu}
  \left(\left(1-\mu^2\right)\hat\Omega\right)
  \nonumber \\
  &&+ \left(\frac{\partial}{\partial\mu}\left(\left(1-\mu^2\right)^2
  \frac{\partial\hat\Omega}{\partial\mu}\right) -
  2\left(1-\mu^2\right)\hat\Omega\right)\frac{\partial V}{\partial\mu}
  \nonumber \\
  &&- \left(\frac{\partial}{\partial\mu}\left(\left(1-\mu^2\right)^2
  \frac{\partial\hat\Omega_\mathrm{A}}{\partial\mu}\right) -
  2\left(1-\mu^2\right)\hat\Omega_\mathrm{A}\right)
  \frac{\partial A}{\partial\mu}
  \nonumber \\
  &&- \left(\hat{\cal L}A\right)\frac{\partial}{\partial\mu}
  \left(\left(1-\mu^2\right)\hat\Omega_\mathrm{A}\right),
  \label{A2}
\end{eqnarray}
the equation for the potential $B$ of toroidal magnetic disturbances
\begin{eqnarray}
   \lefteqn{\hat\omega\left(\hat{\cal L}B\right)=
   - \mathrm{i}\, \frac{\epsilon_\eta}{\hat{\lambda}^2}\left(\hat{\cal L}B\right)
   + m\hat{\cal L}\left(\hat\Omega B\right)
   - m\hat{\cal L}\left(\hat\Omega_\mathrm{A} W\right)}
   \nonumber \\
   &&-\ m^2\frac{\partial\hat\Omega}{\partial\mu} A
   - \frac{\partial}{\partial\mu}\left(
   \left(1-\mu^2\right)^2\frac{\partial\hat\Omega}{\partial\mu}
   \frac{\partial A}{\partial\mu}\right)
   \nonumber \\
   &&+\ m^2\frac{\partial\hat\Omega_\mathrm{A}}{\partial\mu} V
   + \frac{\partial}{\partial\mu}\left(
   \left(1-\mu^2\right)^2\frac{\partial\hat\Omega_\mathrm{A}}{\partial\mu}
   \frac{\partial V}{\partial\mu}\right) ,
   \label{A3}
\end{eqnarray}
the equation for the poloidal magnetic disturbances
\begin{equation}
   \hat\omega\left(\hat{\cal L}A\right) =
   - \mathrm{i}\, \frac{\epsilon_\eta}{\hat{\lambda}^2}\left(\hat{\cal L}A\right)
   + m\hat\Omega\left(\hat{\cal L}A\right)
   - m\hat\Omega_\mathrm{A}\left(\hat{\cal L}V\right) ,
   \label{A4}
\end{equation}
and the equation for the entropy perturbations
\begin{equation}
   \hat\omega S = - \mathrm{i}\ \frac{\epsilon_\chi}{\hat{\lambda}^2}
   S + m\hat\Omega S + \hat{\cal L}V .
   \label{A5}
\end{equation}

The equations include finite diffusion via the parameters
\begin{equation}
  \epsilon_\nu = \frac{\nu N^2}{\Omega_0^3 r^2}, \ \ \
  \epsilon_\eta = \frac{\eta N^2}{\Omega_0^3 r^2}, \ \ \
  \epsilon_\chi = \frac{\chi N^2}{\Omega_0^3 r^2}, \ \ \
  \label{A6}
\end{equation}
where $\nu$, $\eta$, and $\chi$ are the viscosity, the magnetic diffusivity
and the heat conductivity, respectively.

The disturbances in physical units can be recovered with 
\begin{eqnarray}
\lefteqn{   P_\mathrm{v} = \left(\Omega_0 r^2/k\right) V,\ \
   T_\mathrm{v} = \left(\Omega_0 r^2\right) W,\ \
   s' = -\frac{\mathrm{i}C_\mathrm{p} N^2}{g k r} S ,}
   \nonumber \\
\lefteqn{   P_\mathrm{m} = \left(\sqrt{\mu_0\rho}\Omega_0 r^2 / k\right) A,\ \
   T_\mathrm{m} = \left(\sqrt{\mu_0\rho}\Omega_0 r^2 \right) B,} 
   \label{A7}
\end{eqnarray}
and the vector magnetic and velocity fields can then be restored with  
\begin{eqnarray}
\lefteqn{   {\vec B}' = \frac{{\vec e}_r}{r^2}\hat{\cal L}P_\mathrm{m}
  - \frac{{\vec e}_\theta}{r}\left(\frac{1}{\sin\theta}
  \frac{\partial T_\mathrm{m}}{\partial\phi} +
  \frac{\partial^2 P_\mathrm{m}}{\partial r\partial\theta}\right)}
  \nonumber \\
  &&+\ \frac{{\vec e}_\phi}{r}\left(\frac{\partial
  T_\mathrm{m}}{\partial\theta} -
  \frac{1}{\sin\theta}\frac{\partial^2 P_\mathrm{m}}
  {\partial r\partial\phi}\right) ,
  \nonumber \\
\lefteqn{  {\vec u}' = \frac{{\vec e}_r}{r^2}\hat{\cal L}P_\mathrm{v}
  - \frac{{\vec e}_\theta}{r}\left(\frac{1}{\sin\theta}
  \frac{\partial T_\mathrm{v}}{\partial\phi} +
  \frac{\partial^2 P_\mathrm{v}}{\partial r\partial\theta}\right)} 
  \nonumber \\
  &&+\ \frac{{\vec e}_\phi}{r}\left(\frac{\partial
  T_\mathrm{v}}{\partial\theta} -
  \frac{1}{\sin\theta}\frac{\partial^2 P_\mathrm{v}}
  {\partial r\partial\phi}\right),
   \label{A8}
\end{eqnarray}
where $\vec{e}_r$, $\vec{e}_\theta$ and $\vec{e}_\phi$ are unit
vectors in radial, meridional and azimuthal directions, respectively.

\begin{thebibliography}{}
\bibitem[1999]{BCM99}
    Barnes,\,G., Charbonneau,\,P., MacGregor,\,K.B.:
    1999, \apj\ 511, 466
\bibitem[2005]{Bea05}
    Barnes,\ J.R., Cameron,\ A.C., Donati,\ J.-F., James,\ D.J., 
    Marsden,\,S.C., Petit,\,P.: 
    2005, \mnras\ 357, L1 
\bibitem[1997]{BA97}
    Basu,\,S., Antia,\,H.M.:
    1997, \mnras\ 287, 189
\bibitem[2003]{C03}
    Cally,\,P.S.:
    2003, \mnras\ 339, 957
\bibitem[1999a]{CDG99}
    Charbonneau,\,P., Dikpati,\,M., Gilman,\,P.A.:
    1999a: \apj\ 526, 523
\bibitem[1999b]{Cea99}
    Charbonneau,\ P., Christensen-Dalsgaard,\ J., Henning,\ R.,
    Lar\-sen,\ R.M., Schou,\ J., Thompson,\ M.J., Tomczyk,\ S.:
    1999b, \apj\ 527, 445
\bibitem[1991]{C-Dea91}
    Christensen-Dalsgaard,\,J., Gough,\,D.O., Thompson,\,M.J.:
    1991, \apj\ 378, 413
\bibitem[1987]{DK87}
    Dziembowski,\,W., Kosovichev,\,A.G.:
    1987, AcA 37, 341
\bibitem[2001]{G01}
    Garaud,\,P.: 
    2001, \mnras\ 324, 68
\bibitem[2005]{G05}
    Gilman,\,P.A.:
    2005, AN 326, 208
\bibitem[2007]{GDM07}
    Gilman,\,P.A., Dikpati,\,M., Miesch,\,M.S.:
    2007, \apjs\ 170, 203
\bibitem[1999]{KR99}
    Kitchatinov,\,L.L., R\"udiger,\,G.:
    1999, A\&A 344, 911
\bibitem[2006]{KR06}
    Kitchatinov,\,L.L., R\"udiger,\,G.:
    2006, A\&A 453, 329
\bibitem[2007]{KR07}
    Kitchatinov,\,L.L., R\"udiger,\,G.:
    2007, astro-ph/0701847
\bibitem[2001]{KST01}
K\"uker, M., Stix, M.: 2001, A\&A 366, 668
\bibitem[2004]{MG04}
    Miesch,\,M.S., Gilman,\,P.A.:
    2004, SoPh 220, 287
\bibitem[1985]{PT85}
    Pitts,\,E., Tayler,\,R.J.:
    1985, \mnras\ 216, 139
\bibitem[1997]{RK97}
    R\"udiger,\,G., Kitchatinov,\,L.L.:
    1997, AN 318, 273
\bibitem[2007]{RK07}
    R\"udiger,\,G., Kitchatinov,\,L.L.:
    2007, NJPh 9, 302
\bibitem[1998]{Sea98}
    Schou,\,J., Antia,\,H.M., Basu,\,S., et al.:
    1998, \apj\ 505, 390
\bibitem[2002]{S02}
    Spruit,\,H.\,C.:
    2002, A\&A 381, 923
\bibitem[1990]{SS90}
    Stix,\,M., Skaley,\,D.:
    1990, A\&A 232, 234
\bibitem[1973]{T73}
    Tayler,\,R.J.:
    1973, \mnras\ 161, 365
\bibitem[1981]{W81}
    Watson,\,M.: 
    1981, GApFD 16, 285
\bibitem[2007]{ZBM07}
    Zahn,\,J.-P., Brun,\,A.S., Mathis,\,S.:
    2007, A\&A 474, 145
\end{thebibliography}
\end{document}